%% file: GRB221009A_AGILE_arXiv_v03.tex
\documentclass[twocolumn]{aastex631}  

\usepackage[capitalise]{cleveref}
\usepackage{comment}
\usepackage{graphicx}
\usepackage{multirow}

\revised{Figure 5}

\shorttitle{AGILE -- GRB~221009A}
\shortauthors{The AGILE Team}
\graphicspath{{./}{figures/}}

\begin{document}

\title{AGILE gamma-ray detection of the exceptional GRB~221009A}

\input{author_list}



\begin{abstract}
Gamma-ray emission in the MeV--GeV range from explosive cosmic events is of invaluable relevance to understanding physical processes related to the formation of neutron stars and black holes. Here we report on the detection by the AGILE satellite in the MeV--GeV energy range of the remarkable long-duration gamma-ray burst GRB~221009A. The AGILE onboard detectors have good exposure to GRB~221009A during its initial crucial phases. Hard X-ray/MeV emission in the prompt phase lasted hundreds of seconds, with the brightest radiation being emitted between 200 and 300 seconds after the initial trigger. Very intense GeV gamma-ray emission is detected by AGILE in the prompt and early afterglow phase up to 10,000 seconds. Time-resolved spectral analysis shows time-variable MeV-peaked emission simultaneous with intense power-law GeV radiation that persists in the afterglow phase. The coexistence during the prompt phase of very intense MeV emission together with highly nonthermal and hardening GeV radiation is a remarkable feature of GRB~221009A. During the prompt phase, the event shows spectrally different MeV and GeV emissions that are most likely generated by physical mechanisms occurring in different locations. AGILE observations provide crucial flux and spectral gamma-ray information regarding the early phases of GRB~221009A during which emission in the TeV range was reported.
\end{abstract}

\keywords{Gamma-ray astronomy - gamma-ray burst: general}


\section*{Introduction} \label{sec:intro}
\noindent
The AGILE satellite is an Italian Space Agency mission launched in 2007 and dedicated to gamma-ray astrophysics  \citep{tavani_2009_agile_mission, 2019RLSFN..30S.217P}. AGILE is made of four detectors that are sensitive in different energy ranges ranging from hard X-rays (18--60 keV, SuperAGILE detector), MeV energies (0.35--100 MeV, MCAL detector), and gamma-ray GeV energies (0.03--50 GeV, GRID detector). An anticoincidence (AC) system covers the instrument and is capable of detecting hard X-rays in the range 50--200 keV (see Appendix~\ref{sec:agile_mission}). The AGILE-GRID field of view covers about 2.5 sr at any time with very good angular resolution. Since the satellite spins around the satellite--Sun direction, the GRID detector covers about 10 sr for each spinning revolution ($\sim$7 minutes). Any source within the accessible sky region may be exposed by the GRID for time windows of duration of about 150 s for each revolution, with varying off-axis angles with respect to the normal incidence of the AGILE instrument. The MCAL and AC units are omnidirectional detectors with a sensitivity depending on incidence angles. Earth occultations of sky regions may also play an important role in transient source detection.  These features of the AGILE measurements are important elements for the detection of gamma-ray bursts (GRBs). 

\section{GRB~220910A}
On 2022 October 9, AGILE detected very intense hard X-ray and gamma-ray emission lasting hundreds of seconds from a new transient source \citep{2022GCN.32650....1U,2022ATel15662....1P}, initially classified as an X-ray transient \citep[Swift J1913.1+1946,][]{2022GCN.32632....1D}, and subsequently identified as GRB~221009A \citep{2022GCN.32636....1V}. The event was recorded by several satellites during the prompt phase \citep{2022GCN.32632....1D,2022ATel15651....1N, 2022GCN.32636....1V}, and afterglow emission was detected and monitored in the following days \citep{2022GCN.32652....1B, 2022GCN.32662....1K}. Very-high-energy emission in the TeV energy range was reported during the initial phases of the GRB~\citep{2022GCN.32677....1H}. An overview of observations of GRB~221009A was reported by several X-ray/hard X-ray instruments including \textit{Fermi} Gamma-ray Burst Monitor (GBM; \citealt{2023ApJ...952L..42L}), Konus-Wind \citep{Frederiks:2023bxg}, and subsequently Swift in the afterglow phase \citep{2023ApJ...946L..24W}. The redshift of the optical transient is $z = 0.15095 \pm 0.00005$ \citep{2023arXiv230207891M,2022GCN.32648....1D} which corresponds to a distance of $\sim$750 Mpc. In this Letter we adopt the \textit{Fermi}-GBM trigger time in the hard X-ray band on 2022 October 9, $T_0 =$13:16:59.99 UT \citep{lesage_2022}. 

\section{AGILE Observations}
\begin{figure*}[]
    \centering
    \includegraphics[width=\textwidth]{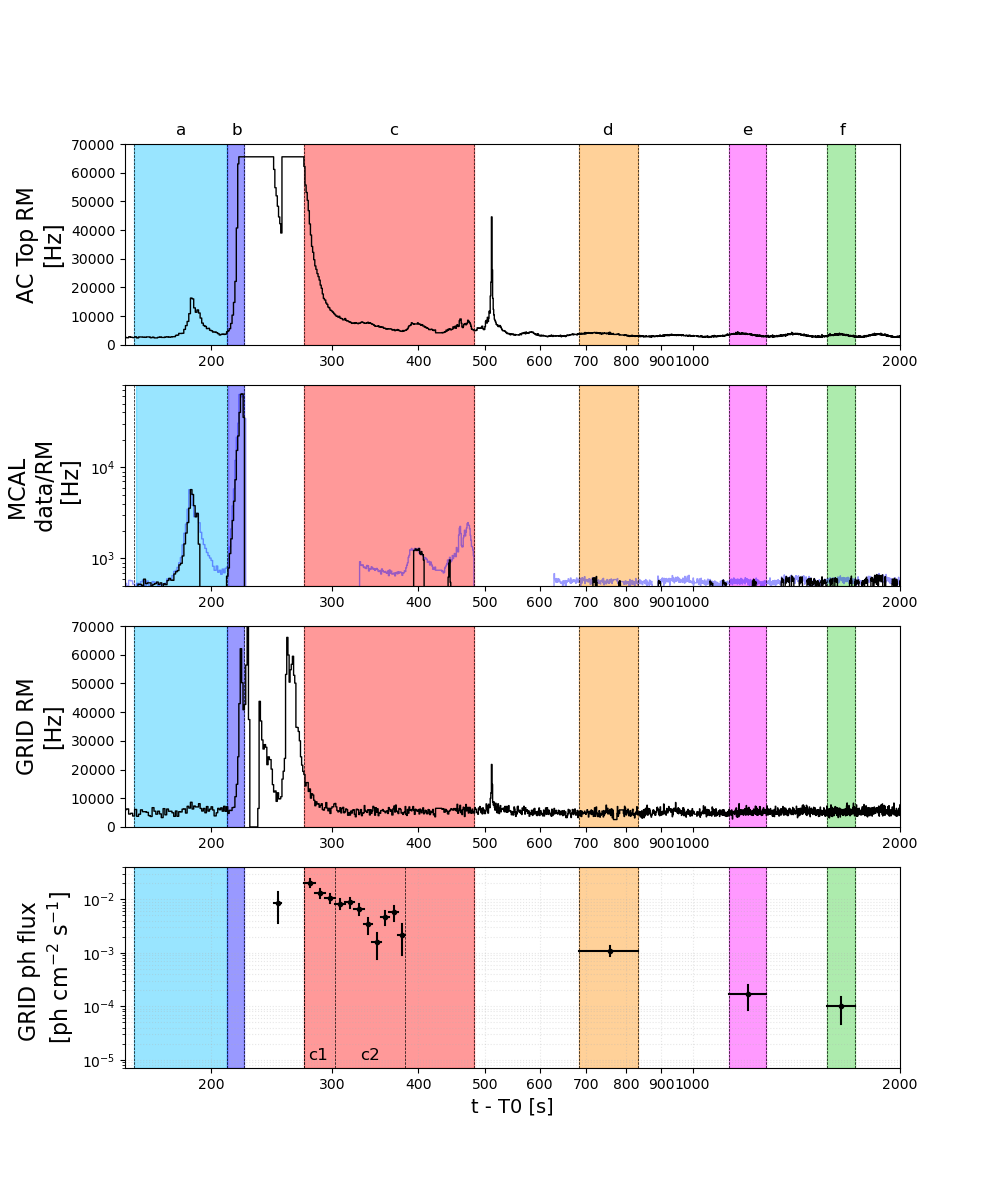}
    \caption{Evolution of the hard X-ray and gamma-ray emission of GRB~221009A as detected by AGILE.
From top to bottom: AC-Top (50--200 keV) RM in linear scale, MCAL (350 keV -- 100 MeV) data (black curve) and RM (blue
curve) in logarithmic scale, ``unvetoed'' GRID RM in linear scale, GRID photon flux lightcurve (50 MeV -- 50 GeV) in logarithmic scale.}
    \label{fig:PLOT1}
\end{figure*}

The GRB~221009A emission is quite complex: the initial triggering event at $T_0$ turns out to be a weak precursor to the brightest part of the GRB~that occurred between $T_0$ + 200 s and $T_0$ + 300 s. The event was recorded by all AGILE detectors active at that time. \Cref{fig:PLOT1} shows the lightcurves of the AC, MCAL and GRID detectors spanning an extended time window of about 2000 s. A detailed view of the first 600 s of the event in the hard X-ray range, as obtained by the AC and MCAL ratemeters (RMs), is reported in Appendix~\ref{sec:agile_observations_grb} (\Cref{fig:PLOTB1}). AGILE detectors recorded the most intense part of the GRB~221109A source activity with no Earth occultations.  Data were gathered with a time modulation of the signal depending on source exposure. 

\begin{figure*}[]
    \centering
    \includegraphics[width=\textwidth]{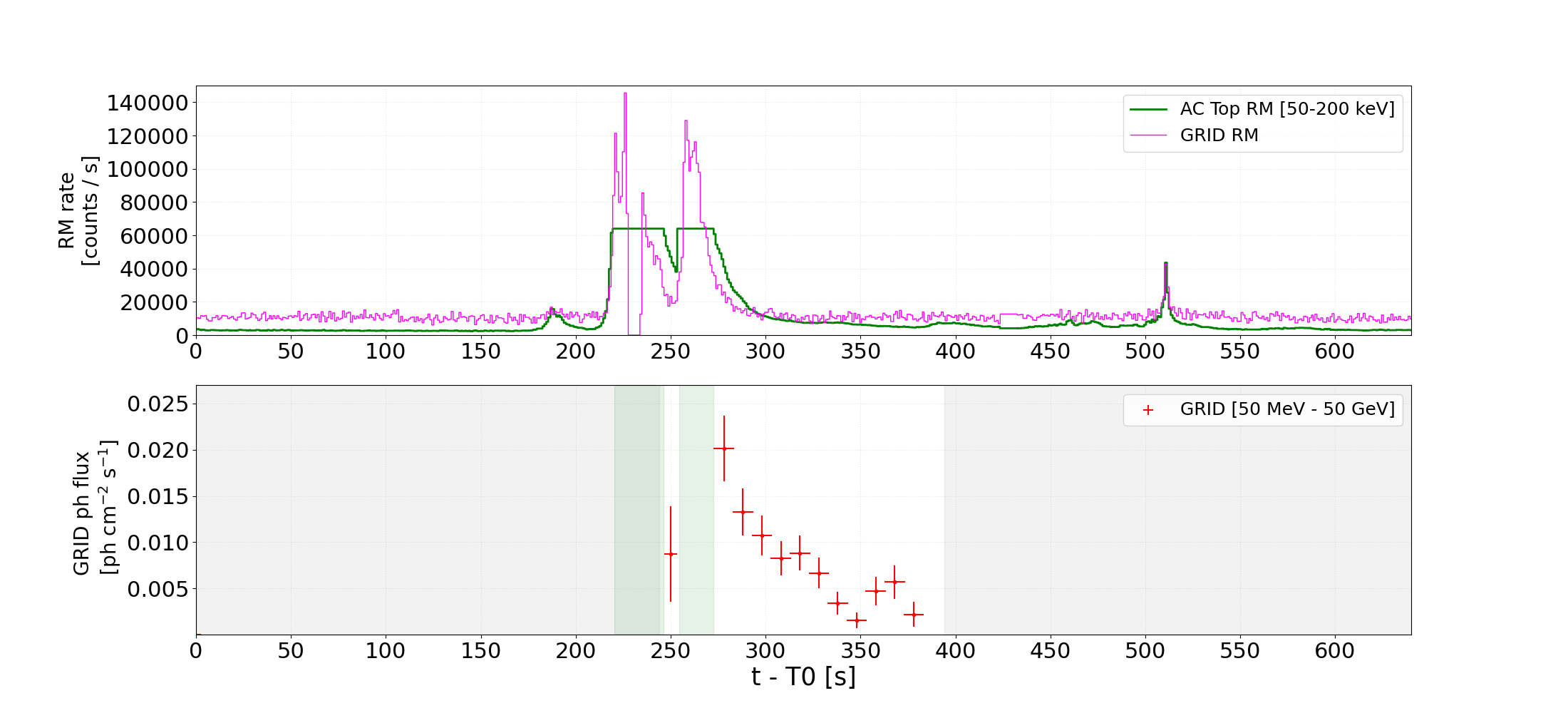}
    \caption{AGILE-GRID lightcurve of the first observation window. Upper panel: AGILE RMs (magenta line: ``unvetoed'' GRID, green line: AC-Top). Lower  panel: AGILE-GRID lightcurve, obtained with ML (10 s time bins for the first OW), for photon energy above 50 MeV. The grey bands represent the time intervals with low GRID exposure, not considered in this analysis (GRB~off-axis angles greater than 60$^\circ$), the green bands show the time intervals of the AC RM saturation, excluded from this analysis.
}
    \label{fig:PLOT2}
\end{figure*}

The hard X-ray emission of the entire prompt phase of GRB~221009A lasted about 500 s and was recorded by the AC, MCAL, and GRID RMs. As shown in \Cref{fig:PLOT1}, this emission was so intense to occasionally saturate the AC, MCAL and GRID ratemeter counters and detectors during the most prominent phase of the emission for 220 s $\lesssim t \lesssim$ 270 s (hereafter, time is measured from $T_0$). The early detection in the hard X-ray range was followed by a remarkably intense gamma-ray emission above 50 MeV revealed by the GRID detector as the GRB~221009A source position entered the GRID field of view (FoV) with an off-axis angle lower than 60$^\circ$. \Cref{fig:PLOT2} (top panel) shows the AC and GRID RM lightcurves in the hard X-rays. \Cref{fig:PLOT2} (bottom panel) shows the gamma-ray emission above 50 MeV as recorded by the GRID detector. Because of a very intense hard X-ray flux, the AC RMs were saturated at the level of making it difficult to properly correct the observed fluxes recorded by the GRID during saturated intervals (the AC is used as charged particle veto in the gamma-ray GRID trigger algorithm; see Appendix~\ref{sec:grid_analysis}). Because of the AC ratemeter saturation, we excluded from the GRID data set the time intervals [220.4, 246.4 s] and [254.4, 272.6 s], marked with green bands in \Cref{fig:PLOT2} (bottom panel). 
However, very bright gamma-ray emission is detected in the first available time bin unaffected by AC saturation (centered at $t$ = 250 s) and, after AC saturation, in the 10 s time interval centered at $t$ = 278 s.

\begin{figure*}[]
    \centering
    \includegraphics[width=\textwidth]{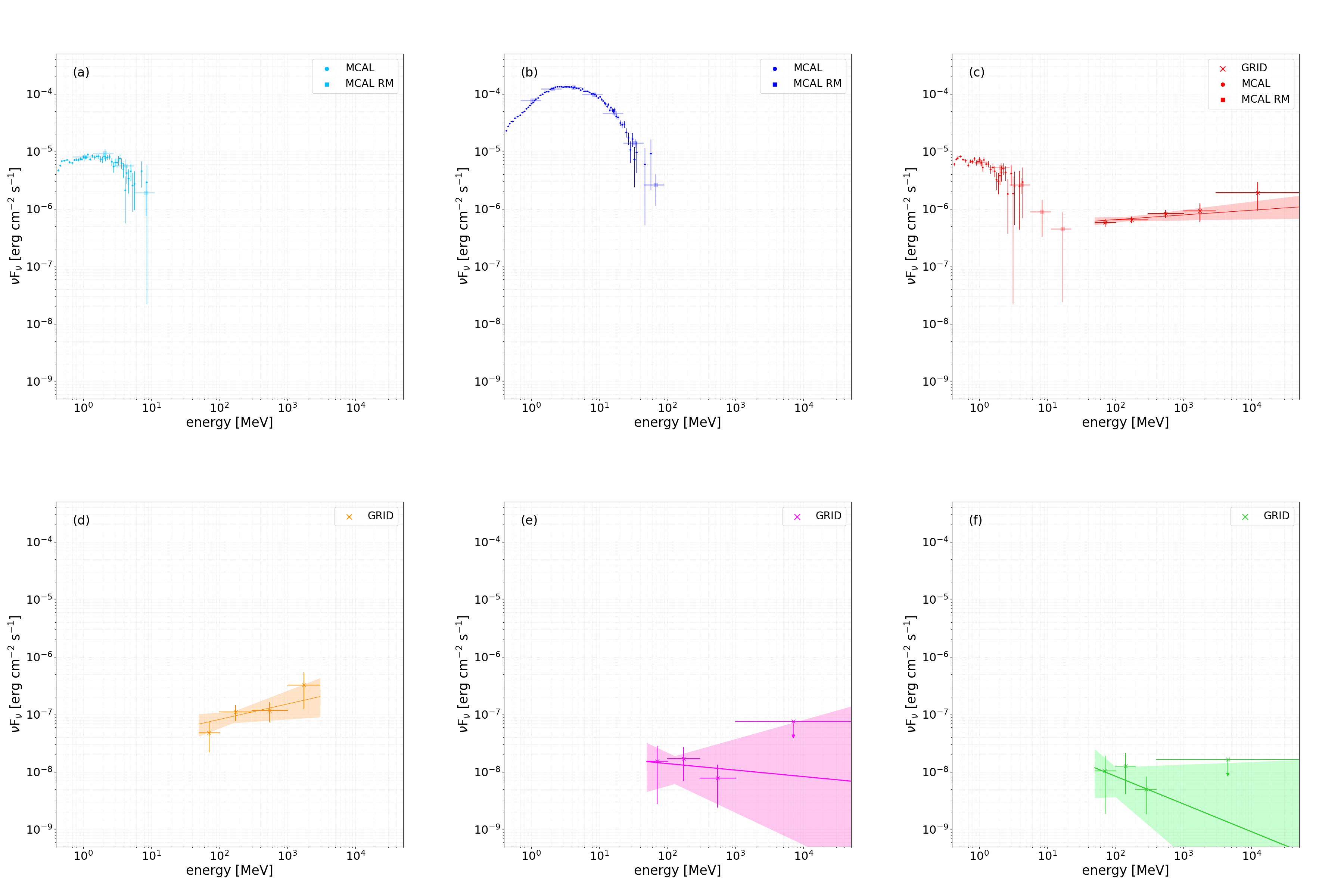}
    \caption{Spectral evolution of the GRB~221009A, as obtained from available data of the AGILE-MCAL and the AGILE-GRID. Panels provide spectral information for the time intervals (a) -- (b) -- (c) -- (d) -- (e) -- (f) defined in \Cref{fig:PLOT1} and \Cref{tab:TABLE1}; the sequence is to be read from left to right, and from top to bottom. The GRID spectra are presented together with the corresponding best-fit curves with their uncertainties (see \Cref{tab:TABLE3}).
}
    \label{fig:PLOT3}
\end{figure*}

It is convenient to define the following time windows in \Cref{fig:PLOT1}. 
\begin{itemize}
\item Interval (a): 155 s $\lesssim t \lesssim$ 211 s, covers a first phase of the intense hard X-ray emission peaking near $t \simeq$ 180 s. Detection by the AC and MCAL RMs. No imaging gamma-ray GRID exposure and no detection by GRID RMs. 
\item Interval (b): 211 s $\lesssim t \lesssim$ 223 s, covers about 10 s of the rapid flux increase to extremely large values that eventually saturated both the AC and MCAL RMs. It is detected by AC, MCAL, and GRID ratemeters with MCAL triggers and spectral information.
\item Interval (c): 273 s $\lesssim t \lesssim$ 482 s, follows the first very intense hard X-ray episode, and it is monitored by the AC and GRID RMs. Starting at $t$ = 273 s very intense gamma-ray emission is detected as the GRID exposure becomes optimal. \Cref{fig:PLOT2} shows the detail of the gamma-ray emission above 50 MeV with time bins of 10 s. The MCAL detector is saturated during the first part of interval (c), and this detector has additional triggers and spectral information around $t$ = 400 s.
\item Interval (d): 684 s $\lesssim t \lesssim$ 834 s, covers a time interval corresponding to the next GRID exposure. The source is detected in gamma-rays in what appears to be the early afterglow phase.
\item Interval (e): 1129 s $\lesssim t \lesssim$ 1279 s, covers the time window corresponding to the third GRID exposure since $T_0$. The source is detected in gamma-rays in the afterglow phase.
\item Interval (f): 1569 s $\lesssim t \lesssim$ 1719 s, covers the time window corresponding to the fourth GRID exposure. The source is detected in gamma-rays in the afterglow phase.
\end{itemize}
\Cref{tab:TABLE1} provides a summary of GRID and MCAL observations selected for this study (in all other cases, the GRID times are rounded to integer values for the sake of simplicity). Several spectral features are worth noticing as summarized in \Cref{fig:PLOT3}.
A high-flux MCAL spectrum in interval (b) is obtained during the very bright rising phase of the prompt emission episode near $t \simeq$ 220 s, showing the brightest spectrum ever detected by AGILE from any previous GRBs. No simultaneous GRID data were available at that time (see \Cref{tab:TABLE1}). 
Nevertheless, a GRID gamma-ray detection with a significance of $\sim$5$\sigma$ is found during the time interval [247, 253 s], just between the two AC saturated windows, with a flux $F = (9 \pm 5) 10^{-3}$ ph cm$^{-2}$ s$^{-1}$  (photon energy between 50 MeV and 50 GeV) and a photon index $\alpha$ = 2.3 $\pm$ 0.6.

Following the second AC saturation interval, very intense gamma-ray emission is detected by the GRID up to $t$ = 383 s with a decreasing flux and average spectral photon index $\alpha$ = 1.92 $\pm$ 0.06 (see \Cref{tab:TABLE3} in Appendix~\ref{sec:grid_analysis}).
Spectral information in the MeV range from MCAL could be obtained near $t$ = 400 s. The coexistence of the MeV and GeV components is shown in the integrated spectrum of interval , which can be considered the beginning of the GRB afterglow (see \Cref{fig:PLOT3}). Interval (d) shows a hardening of the GeV component and an overall decrease of the gamma-ray flux. Spectra of intervals (e) and (f) are suggestive of spectral softening at long timescales as the gamma-ray flux further decreases during the afterglow phase.

\begin{figure*}[]
    \centering
    \includegraphics[width=\textwidth]{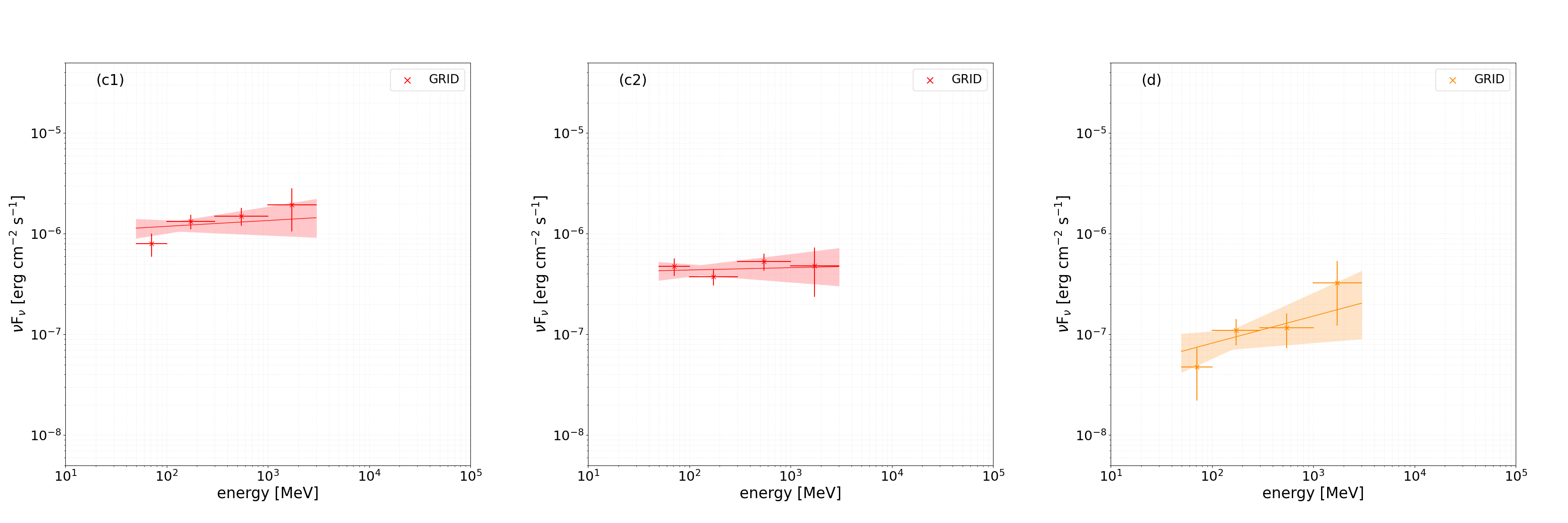}
    \caption{Spectral evolution of the GRB~221009A, as detected by the AGILE-GRID during the afterglow phase detected in gamma rays. Spectral energy distribution (from left to right) for the (c1), (c2), and (d) time intervals. The spectra are presented together with the corresponding best-fit curves with their uncertainties (see \Cref{tab:TABLE4}).
}
    \label{fig:PLOT4}
\end{figure*}

\begin{figure*}[]
    \centering
    \includegraphics[width=\textwidth]{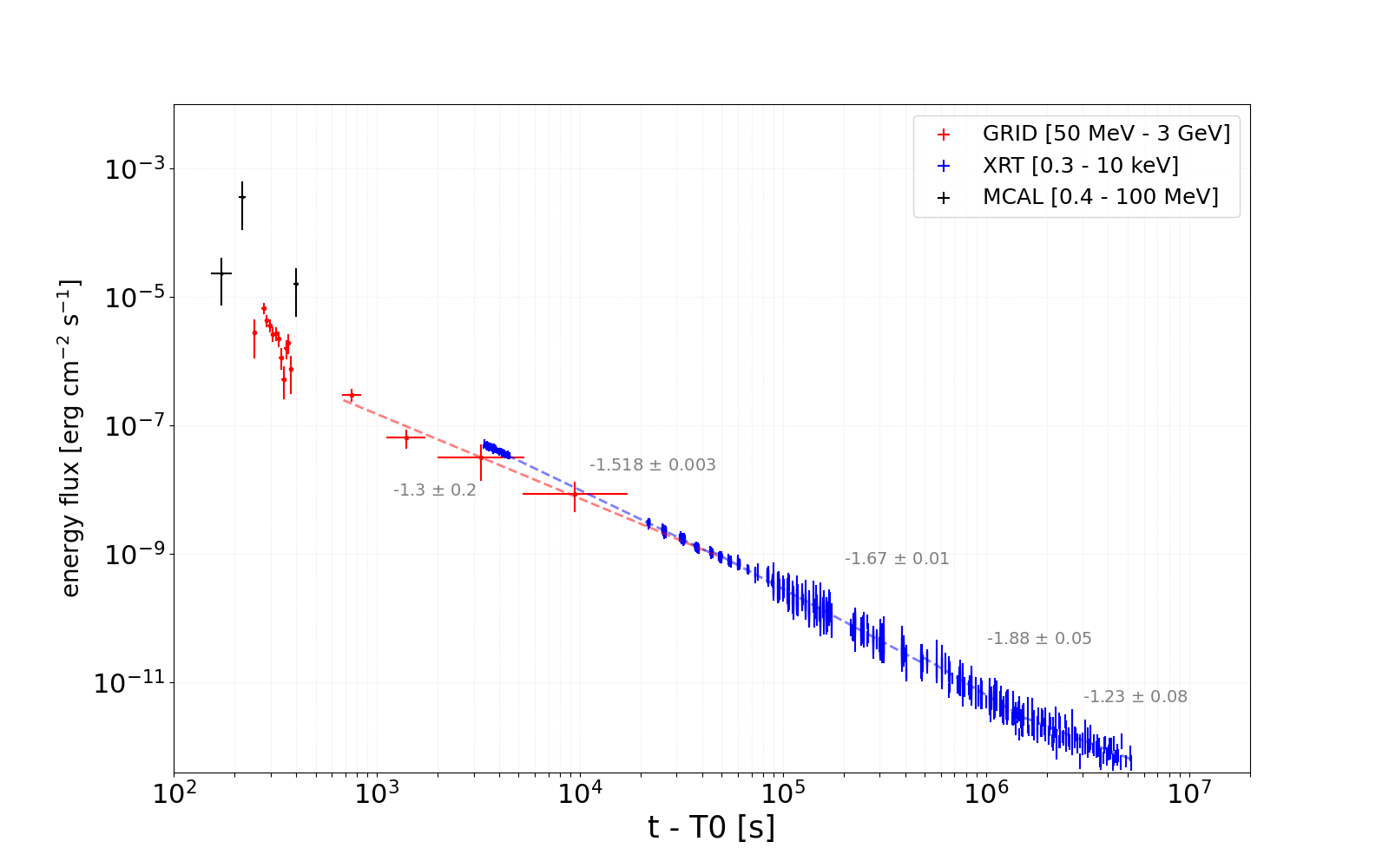}
    \caption{Energy flux evolution of the GRB~221009A: prompt phase and afterglow. Black points: AGILE-MCAL. Red points: AGILE-GRID. Blue points: \textit{Swift}--XRT \citep{2023ApJ...946L..24W}.  Time power-law indices are indicated in the figure.
}
    \label{fig:PLOT5}
\end{figure*}

Spectral hardening in the GeV range as the overall flux decreases in the early phases of the afterglow is made clear in \Cref{fig:PLOT4}, which shows GRID spectra for interval (c), split in two parts, and interval (d). The gamma-ray spectral evolution, as reported by the GRID at the GeV energy range, is suggestive of emission possibly to even larger energies, as indeed reported in \citet{2023ApJ...946L..24W}. It is peculiar to GRB~221009A that the timing of this GeV spectral hardening possibly up to hundreds of GeV energies or beyond occurs in the interval 700--800 s after trigger. It is worth noticing that GRB~190114C, the first GRB~ever announced to have a very-high-energy afterglow in the TeV range \citep{2019Natur.575..455M}, showed the hardening additional GeV spectral component on a timescale of tens of seconds \citep{2020ApJ...904..133U} instead of hundreds of seconds as in the case of GRB~221009A. AGILE observations provide then crucial gamma-ray information for the early phases of GRB~221009A during which intense emission in the TeV  energy range was reported \citep{2022GCN.32677....1H}.
Starting at $t \simeq$ 680 s, the time evolution of the GRID gamma-ray flux F(50 MeV – 3 GeV) can be fit as a power law $F \sim t^{\beta}$ with $\beta = -1.3 \pm 0.2$.  \Cref{fig:PLOT5} shows the gamma-ray flux evolution of GRB~221009A as detected by AGILE as compared to the X-ray afterglow as monitored by \textit{Swift} X-Ray Telescope (XRT; \citealt{2023ApJ...946L..24W}). The temporal power-law agreement between the X-ray and gamma-ray emissions is evident during the afterglow phase.
This latter feature is an important point that relates the afterglow physics of GRB~221009A with other GRBs with prominent gamma-ray emissions (e.g., GRB~130427A, \citealt{2014Sci...343...42A}; GRB~190114C, \citealt{2019Natur.575..455M}; GRB~220101A, \citealt{2022ApJ...933..214U}). Despite the very different timescales and MeV--GeV peak fluxes of the prompt emission, GRB~221009A shares the afterglow dynamics and radiative processes with other GRBs with intense gamma-ray afterglows \citep{2014Sci...343...42A, 2019Natur.575..455M, 2022ApJ...933..214U}.

\begin{table}
    \centering
        \caption{Summary of GRID and MCAL Observations Selected for This Study.}
    \begin{tabular}{|c|c|ccccc|}
    \hline\hline
\multicolumn{2}{|c|}{Intervals} & $t_{\textrm{start}}$ & $t_{\textrm{stop}}$ & MCAL & MCAL & GRID \\
\multicolumn{2}{|c|}{} & & & Data &  RM & Data \\ \hline
\multicolumn{2}{|c|}{} &  (s)   &   (s)  &     &     &  \\ \hline\hline
\multicolumn{2}{|l|}{\multirow{2}{*}{a}}& 154.80 & 192.07 & yes & yes & no \\ 
\multicolumn{2}{|c|}{} & 192.07 & 210.85 & no  & yes & no \\ \hline
\multicolumn{2}{|l|}{b}& 210.85 & 223.23 & yes & yes & no \\ \hline
\multirow{7}{*}{c} & c1 & 273.01 & 303.01 & no & no & yes \\ \cline{2-7}
~ & \multirow{2}{*}{c2} & 303.01 & 329.59 & no & no & yes \\ 
~ & ~  & 329.59 & 383.01 & no & yes & yes \\ \cline{2-7}
~ & \multirow{5}{*}{}  & 383.01 & 393.59 & no & yes & no \\ 
~ & ~ & 393.59 & 407.55 & yes & yes & no \\ 
~ & ~  & 407.55 & 441.99 & no & yes & no \\ 
~ & ~  & 441.99 & 445.11 & yes & yes & no \\ 
~ & ~  & 445.11 & 481.55 & no & yes & no \\ \hline
\multicolumn{2}{|l|}{d} & 684.01 & 834.01 & no & yes & yes \\ \hline
\multicolumn{2}{|l|}{e} & 1129.01 & 1279.01 & no & yes & yes \\ \hline
\multicolumn{2}{|l|}{f} & 1569.01 & 1719.01 & no & yes & yes \\
\hline\hline
    \end{tabular}
\label{tab:TABLE1}
\end{table}

\section{Theoretical modeling}

The relativistic fireball model of the afterglow emission can be applied to an expansion in different environments.  The gas density under different environmental conditions can be represented as $n(r)= A \: r^{-s}$, where we adopt the radial distance $r$, a normalization constant $A$, and the density profile index $s$. Common astrophysical scenarios include an expansion taking place in a constant density medium ($s = 0$), or within a dense stellar environment ($s = 2$). As shown in the case of GRB~190114C \citep{2019Natur.575..455M}, a complete set of multifrequency information regarding the complex afterglow phase is essential for a comprehensive quantitative treatment of GRB~221009A. This is particularly important when distinguishing between the different possible astrophysical scenarios with $s = 0$ or $s = 2$. 

For this reason, we will limit our discussion only to verify that -- in a reasonable scenario of GRB evolution in a constant density medium ($s = 0$) -- the AGILE data presented in this work play a crucial role in order to better define the physics of this extremely bright event. Remarkably, this event has been detected up to very-high-energy $\gamma$-rays by LHAASO \citep{LHAASO_2023}. Thus, in this Letter we model the AGILE data together with the LHAASO data for a simultaneous observing time window. We find that the combined data sets effectively constrain the physical parameter space, being consistent with the one presented in \citet{LHAASO_2023}.
This is especially significant in understanding the transition from synchrotron to synchrotron self-Compton (SSC) emission. However, since the parameter space is affected by degeneracy, the resulting model may result from a different parameter choice.
A comprehensive exploration of the model fully applied to the data will be addressed in an upcoming publication (L. Foffano et al. 2023, in preparation).

For a physical interpretation of the AGILE data, we consider an external shock model describing the adiabatic expansion of a relativistic blast wave  in a constant density medium ($s = 0$).
This model describes the GRB~afterglow emission as due to synchrotron and inverse Compton (IC) radiation produced by relativistic ﬁreballs expanding in the surrounding medium \citep[e.g.,][]{sari_1998, sari_esin_2001}. 
The evolution of the blast waves is described as a function of time $t$ after the initial event occurring at $T^* = T_0$ + 226 s (here we assume, for simplicity, the same reference time adopted in \citealp{LHAASO_2023}). The temporal dependence of the radial distance $r$ in the observer's frame is described by $r = 4 \Gamma^2 c t$. 
The circumburst density of the surrounding medium is described by $n(r) \equiv n_0$, assuming that it is distributed homogeneously over the radial distance. 
 The shock front is expanding with bulk Lorentz factor $\Gamma(r)$, accelerating electrons and positrons over a power-law energy distribution $N(\gamma) = N_0 \, \gamma^{-p}$ above a minimum energy $\gamma > \gamma_m = \frac{p - 2}{p - 1} \frac{m_p}{m_e} \epsilon_e \Gamma$. A homogeneous magnetic field $B =  \Gamma c \,
 \sqrt{32 \pi n(r) m_p \epsilon_B}$ is assumed to be cospatial with the accelerating particles. Here we adopt $\sigma_T$ as the Thomson cross section; $m_e$ and $m_p$ as the electron and proton mass, respectively; $\epsilon_e$ and $\epsilon_b$ as the electron and magnetic field efficiencies, respectively.
 
During the acceleration process, electrons with energy above the cooling Lorentz factor of the electrons $\gamma > \gamma_c = \frac{6 \pi m_e c}{\sigma_T \Gamma B^2 t}$ lose an important portion of their energy via synchrotron cooling within a time $t$.  
Particularly important is the relation between $\gamma_m$ and $\gamma_c$, which correspond to two different physical regimes. When $\gamma_c > \gamma_m$, the particles are in a slow-cooling regime, and only those with $\gamma > \gamma_c$ cool efficiently. Conversely, when $\gamma_c < \gamma_{m}$, the particles are in a fast-cooling regime, and all of them are emitting efficiently. In our model, both regimes are considered depending on the evolution of $\gamma_m$ and $\gamma_c$ over time. 

For the modeling of GRB~221009A, we use the following set of parameters. 
In a structured-jet GRB scenario \citep{sari_1999,dai_2001}, we adopt an exceptional total GRB~isotropic equivalent energy $E_{\mathrm{iso}} = 1.5\cdot 10^{55}$ erg (larger than the upper limit of the $E_{\mathrm{iso}}$ distribution shown in \citealp{atteia_2017}, but consistent with the value reported in \citealp{LHAASO_2023}), an initial shock bulk Lorentz factor $\Gamma_0 =700$, the power-law index of the particle distribution $p=2.08$, electron energy efficiency $\epsilon_e =0.05$, magnetic field efficiency $\epsilon_b = 0.002$, and constant particle density $n_0 = 0.65$ cm$^{-3}$ (consistent with the values reported in \citealp{2019Natur.575..455M} for GRB~190114C, and in \citealp{LHAASO_2023} for GRB~221009A). In this model, we adopt the Thompson cross section or the Klein-Nishina cross section \citep{klein_nishina_1929} depending on the physical conditions and the specific regime of the scattering. Additionally, internal $\gamma\gamma$ absorption suppression is considered depending on the interacting photon energies and the radius of the internal shock. Absorption at gamma-rays by interaction with the extragalactic background light (EBL) is also shown adopting the model by \citet{ebl_dominguez_2011}.

\begin{figure*}[]
    \centering
    \includegraphics[width=0.7\textwidth]{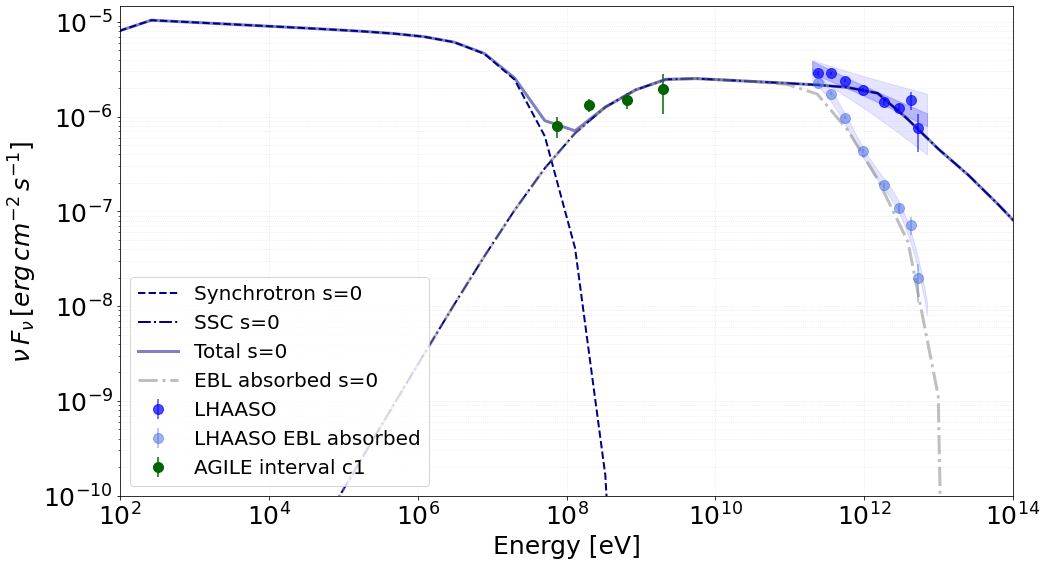}
    \caption{Spectral energy distribution of the external shock model with the set of parameters adopted in the text. Data from the GRID detector of the AGILE instrument for the c1 time interval are reported in green: [273, 303 s] after $T_0$, corresponding to [47, 77 s] after $T^*$ (the reference time adopted in \citealp{LHAASO_2023}). Additionally, we include also data from the LHAASO observatory \citep{LHAASO_2023} for time interval [248, 326 s] after $T_0$, corresponding to [22, 100 s] after $T^*$. LHAASO data points are reported in blue, both for the observed data and the EBL deabsorbed data (faded). More details are reported in the main text.  
}
    \label{fig:SEDAGILELHAASO}
\end{figure*}

In \Cref{fig:SEDAGILELHAASO}, we show the spectral energy distribution of the external shock model with the set of parameters described above between $t_1 = T^*$ + 22 s and $t_2 = T^*$ + 100 s (248 s and 326 s after $T_0$, respectively). Data from the GRID detector of the AGILE instrument are reported. Additionally, we include also data from the LHAASO observatory detected within the aforementioned time window \citep{LHAASO_2023}.
It is interesting to note that the GRID data and LHAASO data are well described by IC emission of the afterglow of GRB~221009A in the considered time interval. 

During that time interval, MCAL data are strongly influenced by prompt emission of the GRB, which is not described by the afterglow model. Additionally, there are no MCAL data strictly simultaneous with GRID and LHAASO observations together. Thus, MCAL data are not presented in the multifrequency spectral energy distribution in \Cref{fig:SEDAGILELHAASO}.

\section{Discussion and conclusions}
The unusual intensity of the high-energy emission of GRB~221009A during the prompt phase can be the result of different factors, including a large (isotropically equivalent) energy available for the burst $E_0 \simeq 10^{55}$ erg \citep{2022GCN.32636....1V}, a large radiative efficiency, a relatively small angle $\theta$ of the jet axis with respect to the line of sight \citep{bright_2023,laskar_2023,negro_2023,LHAASO_2023}, and a relatively small source distance. On the other hand, the X-ray lightcurve of GRB~221009A of the prompt phase shows complex features and a duration that can be attributed to peculiar properties of the central source, or to the fireball expansion in an environment different from that of other similar GRBs. Prompt emission can originate from synchrotron radiation of relativistic electrons/positrons \citep{1994ApJ...430L..93R} or reprocessed radiation from an optically thick photosphere \citep{2009A&A...498..677B}. The MCAL spectrum of interval (b), shown in \Cref{fig:PLOT3}, represents the emission during the very rapid rising portion of the hard X-ray lightcurve. The emission peaks at $E_{\mathrm{peak}} \simeq 3$  MeV and the low-energy spectral index is $\sim$1. Above $E_{\mathrm{peak}}$ the spectral intensity significantly decreases, as it is typical of other GRBs. The (isotropically equivalent) peak luminosity in the MeV band is $L_{\mathrm{MeV}} \simeq 10^{52}$ erg s$^{-1}$.  There is no information from the AGILE-GRID during the interval (b) because of lack of exposure. As the GRID exposure allows it during interval (c), a very prominent and hard gamma-ray emission is produced with a spectrum quite different from the decaying part of the MeV component. As is evident in \Cref{fig:PLOT2}, the  prompt emission is supplemented by an additional GeV component that we attribute to IC emission of high-energy electrons and positrons in a bath of soft photons. Given the spectral features of interval (c), the SSC model can be most likely used in an afterglow-like expansion scenario rather than the physical conditions of the prompt phase \citep{2009A&A...498..677B}. The prompt MeV emission and the GeV SSC component are then manifestations of two different emitting regions \citep[as modeled for GRB~190114C; see e.g., ][]{2019Natur.575..455M}, the former being related to the inner (and probably optically thick region), and the latter being produced in an optically thin and relativistically expanding environment to avoid pair creation photon suppression at GeV as well as TeV energies. We notice that the GRID (isotropically equivalent) peak luminosity and total energy radiated in the GeV band during interval (c) are $L_{\mathrm{GeV}} \simeq 10^{50}$ erg s$^{-1}$ and $E_{\mathrm{GeV}}\simeq 6\times 10^{51}$ erg, respectively.

GRB~221009A, with its remarkable features regarding the intensity, spectral, and duration properties, belongs to a class of GRBs showing the dramatic transition between prompt and afterglow emission with a phase of coexistence of MeV and GeV emissions of very different spectral properties. AGILE data provide crucial information during the most important phases of the GRB~emission in terms of gamma-ray flux intensity and spectral evolution and contribute to the physical modeling that will be undertaken with the complete set of multifrequency data.

\begin{acknowledgments}

The authors thank the editor and the anonymous referee for stimulating comments on the manuscript.

AGILE is a mission of the Italian Space Agency (ASI), with scientific and programmatic participation of Istituto Nazionale di Astrofisica (INAF) and Istituto Nazionale di Fisica Nucleare (INFN). 
This work was carried out in the frame of the Addendum n.6 and n.7 -- Accordo ASI-INAF No. I/028/12/0 for AGILE.

\end{acknowledgments}


\appendix

\section{The AGILE Mission}
\label{sec:agile_mission}

Astrorivelatore Gamma ad Immagini LEggero \citep[AGILE; ][]{tavani_2009_agile_mission} is a space mission of the Italian Space Agency (ASI) devoted to X-ray and gamma-ray astrophysics, developed with scientific and programmatic participation by INAF, INFN, CIFS, several Italian universities and industrial contractors. The AGILE payload consists of the Gamma Ray Imager Detector (GRID) and a hard X-ray detector SuperAGILE for the simultaneous detection and imaging of photons in the 30 MeV--50 GeV and in the 18--60 keV energy ranges, respectively. The payload is completed by two all-sky nonimaging detectors: a mini-calorimeter (MCAL) sensitive in the energy range 350 keV -- 100 MeV, and an anticoincidence (AC) system in the band 50--200 keV.
The AGILE satellite has been operating nominally since 2007 in a low Earth equatorial orbit, and in its spinning observation mode it performs a monitoring of about 80\% of the entire sky with its imaging gamma-ray detector every $\sim$7 minutes. 
The AGILE data are downlinked every $\sim$95 minutes to the ASI Malindi ground station in Kenya, transmitted first to the Telespazio Mission Control Center at Fucino, and then sent within $\sim$5 minutes after the end of each contact to the AGILE Data Center (ADC), which is part of the ASI multimission Space Science Data Center (SSDC, \citealt{2019RLSFN..30S.217P}). The ADC oversees all the scientific-oriented activities related to the analysis and archiving and distribution of the AGILE data. The AGILE ground segment alert system is distributed among ADC and the AGILE Team Institutes, and it combines the ADC quick look with the AGILE Science Alert System developed by the AGILE Team \citep{2019ExA....48..199B}.\\

\section{AGILE observations of GRB~221009A}
\label{sec:agile_observations_grb}

\begin{figure}[]
    \centering
    \includegraphics[width=0.95\textwidth]{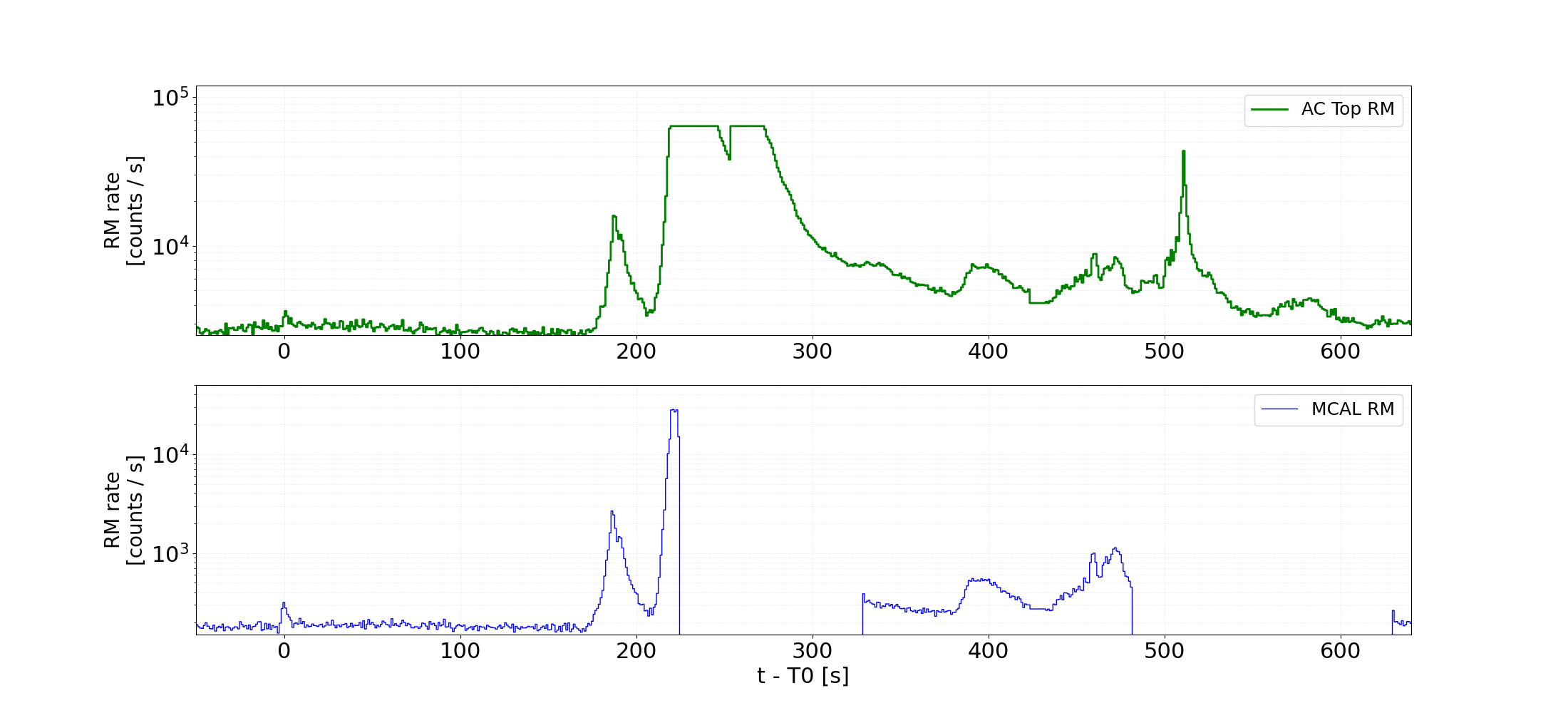}
    \caption{Time evolution of the AGILE scientific RMs during the GRB~221009A. Top panel: AC-Top RM. Bottom panel: MCAL RM.
}
    \label{fig:PLOTB1}
\end{figure}

We analyzed the GRB by taking into account all the available data detected by the instruments on board the AGILE satellite: GRID and MCAL (SuperAGILE was not in observing mode in the considered time window), including the scientific RMs from the GRID, MCAL, and AC system. RMs are low-time-resolution lightcurves with an almost continuous coverage. In \Cref{fig:PLOTB1}, the AC-Top and MCAL RMs for the first 600 s following $T_0$ are displayed, revealing the complex time evolution of the hard X-ray component during the GRB's early phase. Additionally, we chose to present in this work (e.g., in \Cref{fig:PLOT1,fig:PLOT2}) the ``unvetoed'' GRID RMs: they include the integrated signal recorded by the silicon strips in each Si-tracker plane, before the onboard trigger logic for photon acquisition is applied, taking into account the AC-induced veto signal \citep{tavani_2009_agile_mission}. This raw RM, preceding the preliminary onboard $\gamma$-ray event reconstruction and the first background rejection for $\gamma$-ray photon acquisition, is also influenced by the high level of incoming X-rays. 

We performed an analysis of the whole data set, starting from $T_0$ = 2022-10-09 13:16:59.99 UT. \Cref{tab:TABLE1} gives an overview of the AGILE data (MCAL and GRID photon acquisition) within the time interval [150, 1800 s]. Time is measured from $T_0$.

\begin{figure}[]
    \centering
    \includegraphics[width=0.8\textwidth]{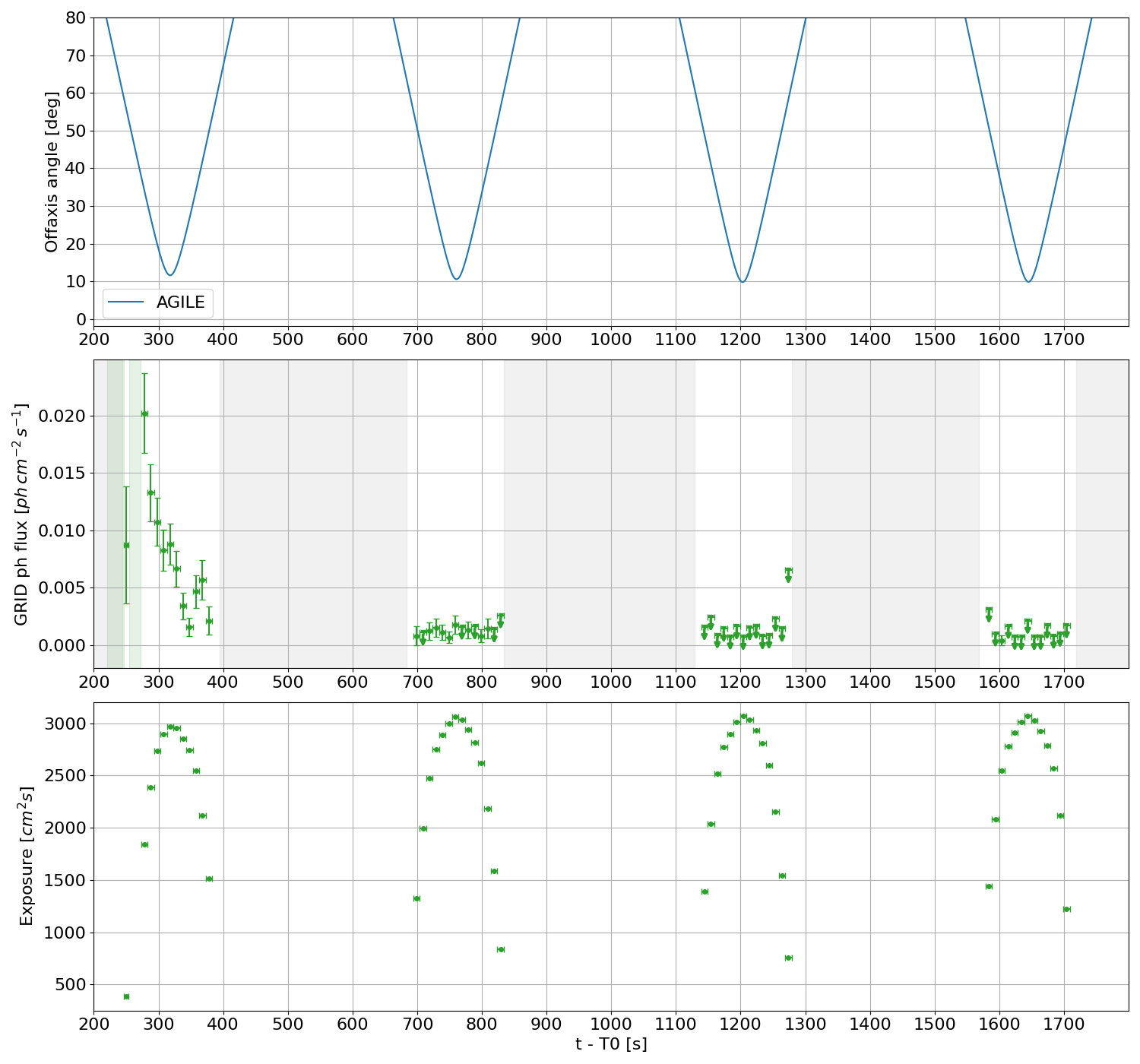}
    \caption{Observational parameters of the GRB~221009A, as detected by the AGILE-GRID. From top to bottom: 
 (1) off-axis angle with respect to the AGILE boresight axis; (2) GRID lightcurve ($E \geq 50$ MeV), the grey bands are related to off-axis angles greater than 60$^\circ$, and the green bands show the time intervals of the AC RM saturation; (3) GRID exposure.
}
    \label{fig:PLOTB2}
\end{figure}

After $T_0$, at $t  \simeq$ 244 s, GRB~221009A entered for the first time in the GRID FoV (off-axis angle lower than 60$^\circ$). Due to the AGILE spinning mode, the GRID exposed the GRB~discontinuously, with $\sim$150 s of source visibility (observation windows, OWs) and $\sim$290 s in which the source is outside the FoV (see \Cref{fig:PLOTB2}). The first four GRID OW time intervals are reported in \Cref{tab:TABLE2}.

\begin{table}
\centering
\caption{GRID Observation Windows (OWs).}
\begin{tabular}{ccc}
\hline
Observation Window & $t_{\text{start}}$  & $t_{\text{stop}}$ \\ 

(OW) & (s) & (s) \\
\hline
OW1 & 244 & 394 \\ 
OW2 & 684 & 834 \\ 
OW3 & 1129 & 1279 \\
OW4 & 1569 & 1719 \\ \hline
\end{tabular}
\label{tab:TABLE2}
\end{table}

The AC-Top scintillator panel became saturated, after the beginning of the prompt phase, by the extraordinarily high incoming X-ray radiation. Due to the onboard veto logic, the AC-Top panel has a prominent role to inhibit the $\gamma$-ray event acquisition in the GRID detector: when the AC-Top scintillator produces a signal due to an incoming radiation (X-ray photon or charged particle), a veto is activated in the GRID onboard logic. This veto signal inhibits the $\gamma$-ray photon acquisition in the GRID detector for a time interval $\Delta t$ = 5.14 $\mu$s.
In order to properly consider this additional dead time for this extremely bright event, we corrected the GRID livetime, by subtracting a $\Delta t$-duration time interval for each corresponding AC-Top count. This correction reduces the effective GRID livetime and, consequently, the source exposure during the prompt phase of the GRB. In this study, we apply this correction to the GRID exposure time, by taking into account the AC-Top count rate only. 
During the time intervals [220.4, 246.4 s] and [254.4, 272.6 s] the RM of the AC-Top panel is saturated (due to a telemetry limit) at a level of 65535 counts s$^{-1}$ (see \Cref{fig:PLOT2}). Thus, we cannot apply an accurate correction to the GRID livetime, since we do not know the exact count rate of the AC-Top panel. For this reason, in order to preserve a conservative analysis approach to this extraordinary event, we have currently excluded these two windows from our GRID analysis. Ongoing studies are in progress to properly quantify the gamma-ray emission detected by the GRID during these phases.\\

\subsection{GRID Analysis}
\label{sec:grid_analysis}

The GRID data have been analyzed with the last available AGILE-GRID software package (Build 25), FM3.119 calibrated filter, H0025 response matrices, and consolidated archive (ASDCSTDk) from the AGILE Data Center at SSDC. Standard data cuts have been applied: South Atlantic Anomaly event cut and 80$^\circ$ Earth albedo filtering have been applied, by taking into account only incoming events with an off-axis angle lower than 60$^\circ$. Event type “G”, for confirmed gamma-ray photon topology -- the most discriminating event selection of the GRID data set -- was used for our analysis. Flux measure and detection significance were calculated by using the AGILE multi-source likelihood analysis (MSLA) software \citep{2012A&A...540A..79B} based on the test statistic (TS) method \citep{1996ApJ...461..396M}. Given the extremely high gamma-ray flux of the GRB, we performed a single source analysis of the region. The GRB~emission was modeled with a simple power law. Galactic and extragalactic diffuse emission parameters were calculated over 2 weeks before $T_0$ and kept fixed in the MSLA. 
We performed the analysis of the GRID data set for the first four OWs, singularly. We excluded the AC-Top saturations windows from OW1. We carried out a spectral analysis over 6 energy bins (0.05--0.10, 0.10--0.30, 0.30--1.00, 1.00--3.00, 3.00--10.00, 10.00--50.00 GeV), by taking into account the instrument response functions (IRFs), including the energy dispersion. For OW2 only, we limited the spectral analysis to a maximum energy of 3 GeV, due to low-statistics issues in the high-energy range. The results are reported in \Cref{tab:TABLE3}.

\begin{table}
    \centering
    \caption{GRID Visibility Intervals.}
    \begin{tabular}{|c|cccccc|}
    \hline
& Time Interval & Energy Range & Detection Significance & Photon Index & Flux & Associated Counts \\ 

& (s, s) & (MeV)  &  & & (ph cm$^{-2}$ s$^{-1}$) & \\ 
\hline
c1 + c2 & [273, 383]   & 50 – 50000 & 46.1$\sigma$ & 1.92 $\pm$ 0.06 & (8.4$\pm$  0.6) $10^{-3}$ & 206 $\pm$  16 \\ 
d & [684, 834]   & 50 – 3000  & 14.4$\sigma$ & 1.7 $\pm$  0.2 & (1.1 $\pm$  0.2) $10^{-3}$ & 53 $\pm$  4 \\ 
e & [1129, 1279] & 50 – 50000 & 5.5$\sigma$ & 2.1 $\pm$ 0.4 & (1.7 $\pm$ 0.8) $10^{-4}$ & 7 $\pm$  3 \\ 
f & [1569, 1719] & 50 – 50000 & 4.5$\sigma$ & 2.5 $\pm$  0.5 & (1.0 $\pm$  0.5) $10^{-4}$ & 6 $\pm$  3 \\ 
\hline
    \end{tabular}
\label{tab:TABLE3}
\end{table}

In order to properly monitor the high-energy spectral evolution of the GRB, during the brightest gamma-ray prompt phase, we further divided OW1 into two subintervals: c1 [273, 303 s] and c2 [303, 383 s]. The resulting spectra are reported in \Cref{tab:TABLE4}.

\begin{table}
    \centering
    \caption{Subintervals c1 and c2.}
    \begin{tabular}{|c|cccccc|}
    \hline
& Subinterval & Energy Range & Detection Significance & Photon Index & Flux & Associated Counts \\ 

& (s, s) & (MeV) &  &  & (ph cm-2 s$^{-1}$) & \\ 
 \hline
c1 & [273, 303] & 50 – 3000 & 32.7$\sigma$ & 1.9 $\pm$ 0.1 & (1.5 $\pm$ 0.2) $10^{-2}$ & 96 $\pm$ 11 \\ 
c2 & [303, 383] & 50 – 3000 & 32.2$\sigma$ & 2.0 $\pm$ 0.1 & (5.4 $\pm$ 0.6) $10^{-3}$ & 107 $\pm$ 12 \\ 
\hline
    \end{tabular}
\label{tab:TABLE4}
\end{table}

Moreover, a 10 s time binning lightcurve was generated for the time interval c (OW1 without the AC-Top saturation windows), in the range 50 MeV -- 50 GeV. In this analysis, the power-law photon index and the position of the source were kept fixed at the values found for the whole c time interval (see \Cref{tab:TABLE3}). The resulting curve is shown in the bottom panel of \Cref{fig:PLOT2}. The last time bin of OW1 has been excluded in order to preserve conservative cuts on the GRID exposure of the source.

To investigate possible delayed gamma-ray emission from the GRB, after OW1, we produced a supplementary GRID lightcurve with time bins of different duration, selected to optimize the photon statistics. In this case, we adopt a standard photon index value of 2.0 for the GRB~spectral power law. We computed the MSLA over a photon energy range 50 MeV -- 3 GeV. The results of this analysis are reported in \Cref{tab:TABLE5} and in \Cref{fig:PLOT5}.

\begin{table}
\centering
\caption{Supplementary GRID Lightcurves with Time Bins of Different Duration, Selected to Optimize the Photon Statistics.}
\begin{tabular}{cccc}
\hline
$t_{\text{start}}$ & $t_{\text{stop}}$  & Detection Significance & $F$ \\

(s]) & (s) &  &  (ph cm$^{-2}$ s$^{-1}$) \\   
\hline
684 & 834 & 15.8$\sigma$ & (9.0 $\pm$ 1.7) $10^{-4}$\\ 
1129 & 1719 & 7.6$\sigma$ & (2.0 $\pm$ 0.6) $10^{-4}$ \\ 
2014 & 5269 & 3.7$\sigma$ & (9.5 $\pm$ 5.2) $10^{-5}$ \\ 
5273 & 16980 & 4.1$\sigma$ & (2.6 $\pm$ 1.2) $10^{-5}$ \\ 
\hline
\end{tabular}
\label{tab:TABLE5}
\end{table}

Integrating the data between 273 s and 383 s, the GRID was also able to locate the GRB~at Galactic coordinates (l, b) = (53.0, 4.3) $\pm~0.1^\circ$ (stat.) $\pm~0.1^\circ$ (syst.) in agreement with the Fermi GBM position \citep{2022GCN.32636....1V}.

\begin{figure}[]
    \centering
    \includegraphics[width=0.6\textwidth]{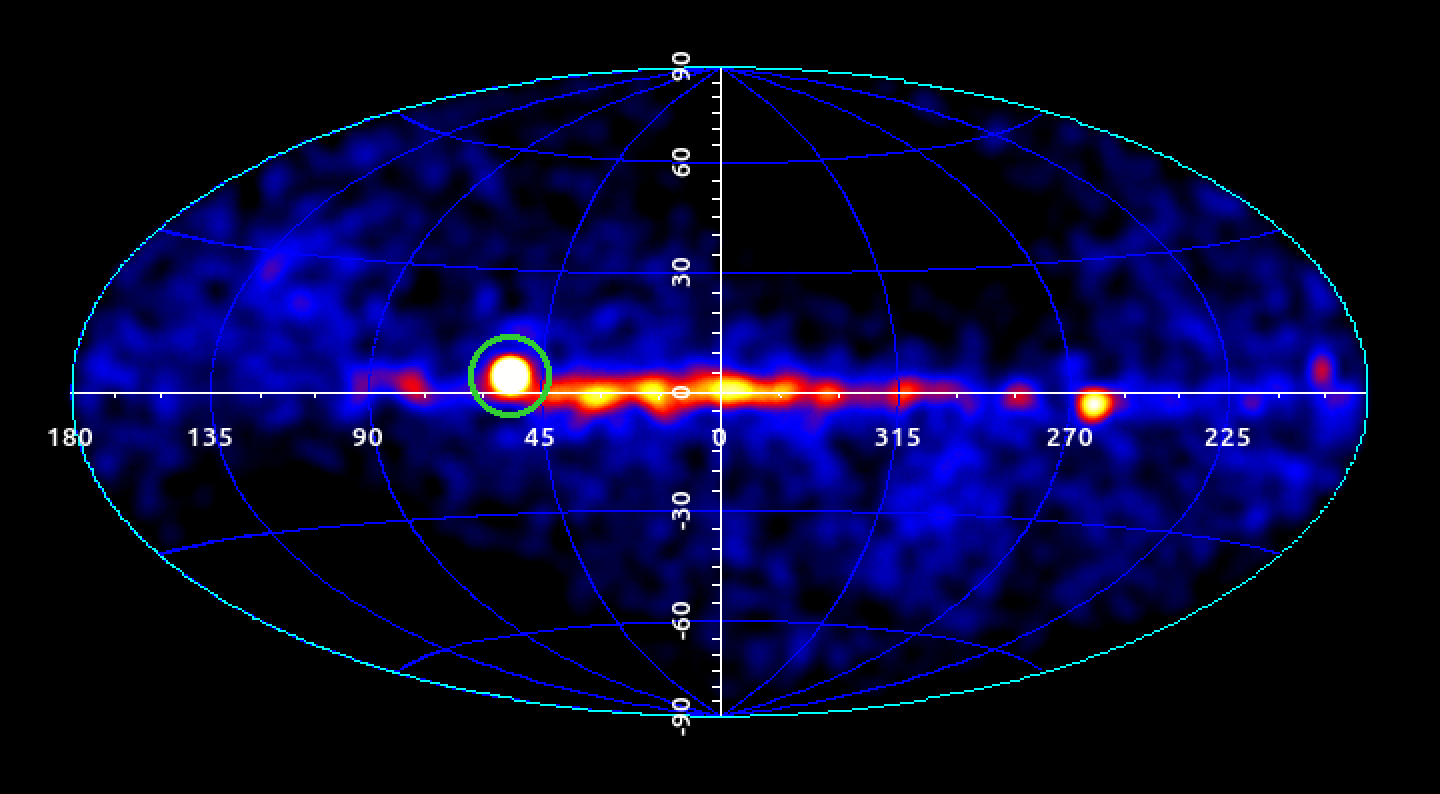}
    \caption{Sky count map above 100 MeV in Galactic coordinates of the AGILE-GRID gamma-ray detector during the time interval [$T_0$, $T_0$ + 48 h]. The gamma-ray source associated with GRB~221009A is shown inside the green circle. The darkened sky regions are due to seasonal lack of exposure of the GRID detector, due to solar panel constraints.
}
    \label{fig:PLOTB3}
\end{figure}

In \Cref{fig:PLOTB3} we present the count map, integrated over 48 h, showing the exceptional luminosity of this event compared with the gamma-ray sky detected by the GRID.

\subsection{MCAL Analysis}
\label{sec:mcal_analysis}

In the time interval between 150 s and 600 s, the AGILE--MCAL was triggered four times, providing partial high-time-resolution photon-by-photon data acquisitions, entirely covering about 66.71 s of the event. The start and end times of each trigger are reported in \Cref{tab:TABLE6}, together with their corresponding duration. On the other hand, the MCAL RM offer a continuous coverage for the whole burst evolution, although with a coarser fixed 1.024 s time resolution. Both MCAL triggered and RM data suffer a lack of data in the time interval between 223.25 s and 391.06 s, due to the extremely high energy release of the event that temporarily blinded the detector. This is particularly evident in \Cref{fig:PLOT1}, where the MCAL lightcurve reaches a count rate above 60 kHz.

\begin{table}
    \centering
    \caption{AGILE MCAL Triggers, with Their Corresponding duration.}
    \begin{tabular}{ccccccc}
    \hline
MCAL Trigger & $t_{\text{start}}$ & $t_{\text{stop}}$ & Trigger Duration \\
&  (s) & (s) &  (s) \\ 

\hline
1 & 154.80 & 192.07 & 37.27 \\ 
2 & 210.85 & 223.23 & 12.38 \\ 
3 & 393.59 & 407.55 & 13.96 \\ 
4 & 441.99 & 445.11 &  3.12 \\ 
\hline
    \end{tabular}
\label{tab:TABLE6}
\end{table}

In order to investigate the evolution of GRB~221009A, we divided the burst in six time intervals (namely, a, b, c, d, e, and f), whose start and end times are reported in \Cref{tab:TABLE1}, together with the available data acquisition status of each AGILE detector.
We performed spectral analysis in the 400 keV - 50 GeV energy range, with the data acquired by the MCAL (both triggered and RM) and GRID detectors. The MCAL spectral analysis was carried out using the XSPEC software package (v 12.12.0; \citealt{1996ASPC..101...17A}).
For what concerns the MCAL triggers, we evaluated the background rate exploiting a triggered onboard acquisition issued before the onset of the GRB~(i.e., from -59.19 s to -44.53 s). The same time interval was adopted also to evaluate the background rate in the MCAL RM data, in order to calibrate and cross-check the obtained values. We divided the background-subtracted counts into the 97 spectral energy channels usually adopted for MCAL spectral analysis and adopted the most suitable response matrix to the attitude of the satellite during each of the six time intervals.\\

\restartappendixnumbering

\nocite{*}
\bibliography{biblio}{}
\bibliographystyle{aasjournal}

\end{document}

%% file: author_list.tex
\correspondingauthor{Giovanni Piano}
\email{giovanni.piano@inaf.it}

\author[0000-0003-2893-1459]{Marco Tavani}
\affiliation{INAF-IAPS Roma, via del Fosso del Cavaliere 100, I-00133 Roma, Italy}
\affiliation{Dipartimento di Fisica, Universit\`a di Roma Tor Vergata, via della Ricerca Scientifica 1, I-00133 Roma, Italy}

\author[0000-0002-9332-5319]{Giovanni Piano}
\affiliation{INAF-IAPS Roma, via del Fosso del Cavaliere 100, I-00133 Roma, Italy}

\author[0000-0001-6347-0649]{Andrea Bulgarelli}
\affiliation{INAF-OAS Bologna, via Gobetti 93/3, I-40129 Bologna, Italy}

\author[0000-0002-0709-9707]{Luca Foffano}
\affiliation{INAF-IAPS Roma, via del Fosso del Cavaliere 100, I-00133 Roma, Italy}

\author[0000-0002-7253-9721]{Alessandro Ursi}
\affiliation{Agenzia Spaziale Italiana (ASI), via del Politecnico snc, I-00133 Roma, Italy}

\author[0000-0003-3455-5082]{Francesco Verrecchia}
\affiliation{ASI Space Science Data Center, SSDC/ASI, via del Politecnico snc, I-00133 Roma, Italy}
\affiliation{INAF - Osservatorio Astronomico di Roma, I-00078 Monte Porzio Catone, Italy}

\author[0000-0001-6661-9779]{Carlotta Pittori}
\affiliation{ASI Space Science Data Center, SSDC/ASI, via del Politecnico snc, I-00133 Roma, Italy}
\affiliation{INAF - Osservatorio Astronomico di Roma, I-00078 Monte Porzio Catone, Italy}

\author[0000-0001-8100-0579]{Claudio Casentini}
\affiliation{INAF-IAPS Roma, via del Fosso del Cavaliere 100, I-00133 Roma, Italy}

\author{Andrea Giuliani}
\affiliation{INAF-IASF Milano, via A. Corti 12, I-20133 Milano, Italy}

\author[0000-0003-2501-2270]{Francesco Longo}
\affiliation{Dipartimento di Fisica, Universit\`a di Trieste, via A. Valerio 2, I-34127 Trieste, Italy}
\affiliation{INFN Trieste, Padriciano 99, I-34012 Trieste, Italy}

\author[0000-0002-3410-8613]{Gabriele Panebianco}
\affiliation{Dipartimento di Fisica e Astronomia, Universit\`a di Bologna, via Gobetti 93/2, I-40129 Bologna, Italy}
\affiliation{INAF-OAS Bologna, via Gobetti 93/3, I-40129 Bologna, Italy}

\author[0000-0002-9894-7491]{Ambra Di Piano}
\affiliation{INAF-OAS Bologna, via Gobetti 93/3, I-40129 Bologna, Italy}

\author{Leonardo Baroncelli}
\affiliation{INAF-OAS Bologna, via Gobetti 93/3, I-40129 Bologna, Italy}

\author[0000-0002-6082-5384]{Valentina Fioretti}
\affiliation{INAF-OAS Bologna, via Gobetti 93/3, I-40129 Bologna, Italy}

\author[0000-0002-4535-5329]{Nicol\`{o} Parmiggiani}
\affiliation{INAF-OAS Bologna, via Gobetti 93/3, I-40129 Bologna, Italy}

\author{Andrea Argan}
\affiliation{INAF-IAPS Roma, via del Fosso del Cavaliere 100, I-00133 Roma, Italy}

\author[0000-0002-3180-6002]{Alessio Trois}
\affiliation{INAF - Osservatorio Astronomico di Cagliari, via della Scienza 5, I-09047 Selargius, Italy}

\author[0000-0003-1163-1396]{Stefano Vercellone}
\affiliation{INAF - Osservatorio Astronomico di Brera, via E. Bianchi 46, I-23807 Merate, Italy}

\author[0000-0001-8877-3996]{Martina Cardillo}
\affiliation{INAF-IAPS Roma, via del Fosso del Cavaliere 100, I-00133 Roma, Italy}

\author[0000-0002-5037-9034]{Lucio Angelo Antonelli}
\affiliation{INAF - Osservatorio Astronomico di Roma, I-00078 Monte Porzio Catone, Italy}

\author{Guido Barbiellini}
\affiliation{Dipartimento di Fisica, Universit\`a di Trieste, via A. Valerio 2, I-34127 Trieste, Italy}
\affiliation{INFN Trieste, Padriciano 99, I-34012 Trieste, Italy}

\author[0000-0003-2478-8018]{Patrizia Caraveo}
\affiliation{INAF-IASF Milano, via A. Corti 12, I-20133 Milano, Italy}

\author[0000-0001-6877-6882]{Paolo W. Cattaneo}
\affiliation{INFN Pavia, via A. Bassi 6, I-27100 Pavia, Italy}

\author{Andrew W. Chen}
\affiliation{School of Physics, Wits University, Johannesburg, South Africa}

\author[0000-0003-4925-8523]{Enrico Costa}
\affiliation{INAF-IAPS Roma, via del Fosso del Cavaliere 100, I-00133 Roma, Italy}

\author[0000-0002-3013-6334]{Ettore Del Monte}
\affiliation{INAF-IAPS Roma, via del Fosso del Cavaliere 100, I-00133 Roma, Italy}

\author{Guido Di Cocco}
\affiliation{INAF-OAS Bologna, via Gobetti 93/3, I-40129 Bologna, Italy}

\author[0000-0002-4700-4549]{Immacolata Donnarumma}
\affiliation{Agenzia Spaziale Italiana (ASI), via del Politecnico snc, I-00133 Roma, Italy}

\author{Yuri Evangelista}
\affiliation{INAF-IAPS Roma, via del Fosso del Cavaliere 100, I-00133 Roma, Italy}

\author{Marco Feroci}
\affiliation{INAF-IAPS Roma, via del Fosso del Cavaliere 100, I-00133 Roma, Italy}

\author[0000-0003-4666-119X]{Fulvio Gianotti}
\affiliation{INAF-OAS Bologna, via Gobetti 93/3, I-40129 Bologna, Italy}

\author[0000-0002-5086-3619]{Claudio Labanti}
\affiliation{INAF-OAS Bologna, via Gobetti 93/3, I-40129 Bologna, Italy}

\author[0000-0003-4871-4072]{Francesco Lazzarotto}
\affiliation{INAF-IAPS Roma, via del Fosso del Cavaliere 100, I-00133 Roma, Italy}

\author{Paolo Lipari}
\affiliation{INFN Roma 1, piazzale Aldo Moro 2, I-00185 Roma, Italy}
\affiliation{Dipartimento di Fisica, Universit\`a La Sapienza, piazzale Aldo Moro 2, I-00185 Roma, Italy}

\author[0000-0002-6311-764X]{Fabrizio Lucarelli}
\affiliation{ASI Space Science Data Center, SSDC/ASI, via del Politecnico snc, I-00133 Roma, Italy}
\affiliation{INAF - Osservatorio Astronomico di Roma, I-00078 Monte Porzio Catone, Italy}

\author[0000-0002-4000-3789]{Martino Marisaldi}
\affiliation{Birkeland Centre for Space Science, Department of Physics and Technology, University of Bergen, Bergen, Norway}
\affiliation{INAF-OAS Bologna, via Gobetti 93/3, I-40129 Bologna, Italy}

\author[0000-0003-3259-7801]{Sandro Mereghetti}
\affiliation{INAF-IASF Milano, via A. Corti 12, I-20133 Milano, Italy}

\author[0000-0002-7704-9553]{Aldo Morselli}
\affiliation{INFN Roma Tor Vergata, via della Ricerca Scientifica 1, I-00133 Roma, Italy}

\author[0000-0001-6897-5996]{Luigi Pacciani}
\affiliation{INAF-IAPS Roma, via del Fosso del Cavaliere 100, I-00133 Roma, Italy}

\author[0000-0002-4590-0040]{Alberto Pellizzoni}
\affiliation{INAF - Osservatorio Astronomico di Cagliari, via della Scienza 5, I-09047 Selargius, Italy}

\author{Francesco Perotti}
\affiliation{INAF-IASF Milano, via A. Corti 12, I-20133 Milano, Italy}

\author[0000-0002-7986-3321]{Piergiorgio Picozza}
\affiliation{Dipartimento di Fisica, Universit\`a di Roma Tor Vergata, via della Ricerca Scientifica 1, I-00133 Roma, Italy}
\affiliation{INFN Roma Tor Vergata, via della Ricerca Scientifica 1, I-00133 Roma, Italy}

\author[0000-0001-7397-8091]{Maura Pilia}
\affiliation{INAF - Osservatorio Astronomico di Cagliari, via della Scienza 5, I-09047 Selargius, Italy}

\author{Massimo Rapisarda}
\affiliation{INAF-IAPS Roma, via del Fosso del Cavaliere 100, I-00133 Roma, Italy}

\author[0000-0001-9702-7645]{Andrea Rappoldi}
\affiliation{INFN Pavia, via A. Bassi 6, I-27100 Pavia, Italy}

\author{Alda Rubini}
\affiliation{INAF-IAPS Roma, via del Fosso del Cavaliere 100, I-00133 Roma, Italy}

\author[0000-0002-7781-4104]{Paolo Soffitta}
\affiliation{INAF-IAPS Roma, via del Fosso del Cavaliere 100, I-00133 Roma, Italy}

\author[0000-0002-2505-3630]{Massimo Trifoglio}
\affiliation{INAF-OAS Bologna, via Gobetti 93/3, I-40129 Bologna, Italy}

\author[0000-0002-1208-8818]{Valerio Vittorini}
\affiliation{INAF-IAPS Roma, via del Fosso del Cavaliere 100, I-00133 Roma, Italy}

\author{Fabio D'Amico}
\affiliation{Agenzia Spaziale Italiana (ASI), via del Politecnico snc, I-00133 Roma, Italy}